\documentclass[12pt]{article}
\usepackage{amsfonts}
\usepackage{bbm,dsfont}
\newcommand{\be}{\begin{equation}}
\newcommand{\ee}{\end{equation}}
\newcommand{\bea}{\begin{eqnarray}}
\newcommand{\eea}{\end{eqnarray}}
\newcommand{\bean}{\begin{eqnarray*}}
\newcommand{\eean}{\end{eqnarray*}}
\font\upright=cmu10 scaled\magstep1
\font\sans=cmss12

\newcommand{\ssf}{\sans}
\newcommand{\stroke}{\vrule height8pt width0.4pt depth-0.1pt}
\newcommand{\Z}{\hbox{\upright\rlap{\ssf Z}\kern 2.7pt {\ssf Z}}}

\newcommand{\C}{{\rlap{\rlap{C}\kern 3.8pt\stroke}\phantom{C}}}
\newcommand{\R}{\hbox{\upright\rlap{I}\kern 1.7pt R}}
\newcommand{\CP}{\C{\upright\rlap{I}\kern 1.7pt P}}

\begin{document}
\pagestyle{plain}

\title{\vskip -70pt
\begin{flushright}
{\normalsize DAMTP-2004-60} \\
\end{flushright}
\vskip 20pt
{\bf The K\"ahler Potential of Abelian Higgs Vortices} \vskip 20pt}

\author{Heng-Yu Chen\thanks{email H.Y.Chen@damtp.cam.ac.uk} \\[10pt]
and \\[10pt]
N. S. Manton\thanks{email N.S.Manton@damtp.cam.ac.uk} \\[15pt]
{\sl Department of Applied Mathematics and Theoretical Physics} \\[5pt]
{\sl University of Cambridge} \\[5pt]
{\sl Wilberforce Road, Cambridge CB3 0WA, UK} \\[15pt]}
\date{July 2004}
\maketitle

\begin{abstract}
We calculate the K\"ahler potential for the Samols metric on the moduli space of Abelian Higgs vortices 
on $\mathbbm{R}^{2}$, in two different ways. 
The first uses a scaling argument. The second is related to the Polyakov conjecture in Liouville field theory.
The K\"ahler potential on the moduli space of vortices on $\mathbbm{H}^{2}$ is also derived, 
and we are led to a geometrical reinterpretation of these vortices. 
Finally, we attempt to find the K\"ahler potential for vortices on $\mathbbm{R}^{2}$ in a third way by relating the vortices to
SU(2) Yang-Mills instantons on $\mathbbm{R}^{2}\times S^{2}$. This approach does not give the correct result, and we
offer a possible explanation for this.  

\end{abstract}
\newpage
\section{Introduction}
\setcounter{equation}{0}
Vortex solutions are known to exist in the (2+1)-dimensional Abelian Higgs model. They 
are the static field configurations minimizing the energy functional.
The Lagrangian density for this model is 
\be
{\mathcal{L}}= 
-\frac{1}{4}F^{\mu\nu}F_{\mu\nu}+\frac{1}{2}\overline{\mathrm{D}_{\mu}\Phi}
\mathrm{D}^{\mu}\Phi-\frac{\lambda}{8}(|\Phi |^{2}-1)^{2}\,.
\ee  
When the coupling constant $\lambda$ takes the critical value of 1, 
there are no net forces between the vortices.
There then exist static configurations satisfying 
the first order Bogomolny equations (see (2.3), (2.4) below).
The $N$-vortex solutions in $\mathbbm{R}^ {2}$ or equivalently the complex plane $\mathbbm{C}$
can be uniquely characterized by where the Higgs field $\Phi$ vanishes \cite{JT}. 
The $N$ unordered Higgs zero locations in $\mathbbm{C}$ are therefore the natural coordinates 
parameterising the space of static $N$-vortex solutions. This space is called the 
moduli space for $N$ vortices, and we shall denote it by ${\mathcal{M}}_{N}$.
These coordinates on ${\mathcal{M}}_{N}$ are called ``collective coordinates''.
${\mathcal{M}}_{N}$ has a natural K\"ahler structure inherited from the kinetic terms of the Lagrangian.

The so-called ``moduli space approximation'' is a powerful approach for studying the 
low energy dynamics of solitonic objects in field theories \cite{M1}. The idea is that, in the low 
energy limit, most of the field degrees of freedom are effectively frozen. The solitonic
dynamics can thus be described by the dynamics in a reduced, finite-dimensional     
space of collective coordinates, which is the moduli space. 
For $N$ vortices, the potential energy is at the absolute minimum everywhere
on the space ${\mathcal{M}}_{N}$ equipped with its K\"ahler metric. If vortices  move
at small velocity, they are trapped close to ${\mathcal{M}}_{N}$, 
and the kinetic energy term of the reduced dynamics dominates. 
The trajectories are the geodesics 
on ${\mathcal{M}}_{N}$, and the scattering of the vortices can be accurately modelled by such    
geodesic motions. 

Finding the K\"ahler metric of ${\mathcal{M}}_{N}$ 
has been the central problem in understanding vortex dynamics
within the geodesic approximation.
A general, but not explicit formula for the metric was first derived by Samols \cite{S}.
Recently, an explicit formula in terms of modified Bessel functions was given for 
the K\"ahler metric on the moduli space of $N$ well separated vortices \cite{MS}. 
A formula for the K\"ahler potential was also given.
The main purpose of this paper therefore is to construct a K\"ahler potential for the more general Samols metric.

The K\"ahler potential and the K\"ahler metric on a complex manifold are related via
\be
g_{rs}=\frac{\partial^{2} \mathcal{K}}{\partial Z_{r}\partial \bar{Z}_{s}}\,,
\ee
where $Z_{r}$ $(\bar{Z_{s}})$, $g_{rs}$ and $\mathcal{K}$ are the  
holomorphic (anti-holomorphic) coordinates, 
K\"ahler metric tensor and K\"ahler potential, respectively. 
Notice that if we add a holomorphic or an anti-holomorphic 
function to $\mathcal{K}$, it still gives the same K\"ahler metric, 
so the modified K\"ahler potential is geometrically equivalent to the 
original one. 
This property becomes important if we want to remove undesirable singularities from the K\"ahler potential. 

In this paper, we present three different approaches to calculating the $N$-vortex K\"ahler potential. 
The first approach is to explicitly construct it from the quantities in the Samols metric.
This is an example of the $\bar{\partial}$-problem. 
We show that, for vortices on $\mathbbm{C}$, it can 
be solved by a scaling argument.

The second approach is inspired by a conjecture of Polyakov \cite{P} relating the so-called 
``accessory parameters'' in the context of uniformization of Riemann surfaces to the regularized Liouville action. 
An accessible mathematical proof was given by Takhtajan and Zograf
based on their earlier work in \cite{TZ}. 
Proofs for the Polyakov conjecture on Riemann surfaces 
with a range of singularities can be found in \cite{CMS}, \cite{HJ}. 
It turns out that in the vortex situation, there are analogous quantities to the accessory parameters, 
and we can construct a modified regularized Liouville action as the generating function for these quantities, which acts as the 
interacting part of the K\"ahler potential for the moduli space $\mathcal{M}_{N}$.

Our third approach was motivated by considering $SU(2)$ Yang-Mills 
instantons on a K\"ahler 4-manifold, the K\"ahler potential of whose moduli space was given by Maciocia \cite{AM}.
The appropriate 4-manifold here is $\mathbbm{R}^{2}\times S^{2}$.
Dimensional reduction of the instantons over $\mathbbm{R}^{2} \times S^{2}$, 
using the $SO(3)$ symmetry of $S^{2}$, results in  
Abelian Higgs vortices over $\mathbbm{R}^{2}$. 
Maciocia's formula suggests a way of obtaining the K\"ahler potential for 
the vortices over $\mathbbm{R}^{2}$ by relating it to the K\"ahler potential of the instantons.  
A promising result is derived, but it appears to be incorrect. 
We discuss this difficulty and its possible resolution in section 6.

An interesting variation of the vortices on $\mathbbm{R}^{2}$ 
is the Abelian Higgs model for vortices defined on the hyperbolic plane $\mathbbm{H}^{2}$ with 
constant Ricci scalar $-1$. Such vortices were shown to be integrable by Witten \cite{W}, 
as the Bogomolny equations in this case can be reduced to the 
Liouville equation. 
The metric on the moduli space of vortices on $\mathbbm{H}^{2}$ was first derived by Strachan \cite{IS}.
In this paper we construct a K\"ahler potential for the metric 
and discuss the geometrical interpretation of the Higgs field and K\"ahler metric.

This paper is organized as follows: In section 2, 
we shall briefly review Abelian Higgs vortices, and Samols' metric on the moduli space of $N$-vortex solutions.
In section 3, the relevant $\bar{\partial}$ problem is described, 
and a general formula for the K\"ahler potential based on a scaling argument will be presented.
It is tested for the case of two well separated vortices. 
In section 4, we shall briefly review the regularized action of Liouville field theory and the Polyakov conjecture,
and show how the interacting part of the K\"ahler potential can be constructed for vortices 
on $\mathbbm{C}$ from an analogous regularized action.
In section 5, we shall discuss vortices in the hyperbolic plane,
present a geometric interpretation for the Higgs field, 
and show that in this case the regularized action is the entire
K\"ahler potential.
In section 6, dimensional reduction of the instantons over $\mathbbm{R}^{2}\times S^{2}$ is presented. Maciocia's 
formula and its relation to the vortex K\"ahler potential are discussed. 
A possible explanation for the discrepancy in applying Maciocia's formula to the 
two-dimensional system will also be given.

\section{Abelian Higgs Vortices}
\setcounter{equation}{0}
We shall be working in the $A_{0}=0$ gauge and at critical coupling $\lambda=1$. The total energy in the static situation is
\be
E=\frac{1}{2}\int d^{2}x\, \left\{F_{12}^{2}+\overline{D_{i}\Phi}D_{i}\Phi+\frac{1}{4}(|\Phi |^{2}-1)^{2}\right\}\,,
\ee
where $F_{12}=\partial_{1}A_{2}-\partial_{2}A_{1}=B$ is the magnetic field 
in the plane and $D_{j}\Phi=\partial_{j}\Phi-iA_{j}\Phi$, $j=1,2$, is the 
covariant derivative of the complex Higgs field $\Phi$. 
The boundary condition for  
$\Phi$ is that $|\Phi|\to 1$ as $|\bf{x}|\to \infty$, so $\Phi$
becomes pure phase and finite energy implies that the gauge field becomes pure gauge at spatial infinity, 
such that $D_{j}\Phi$ vanishes.
The winding number of $\Phi$ at infinity is denoted by $N$, and is assumed to be a positive integer.

We can rearrange $E$ into the Bogomolny form using the standard trick of completing the square
\be
E=\frac{1}{2}\int d^{2}x\, 
\left\{\left(F_{12}+\frac{1}{2}(|\Phi |^{2}-1)\right)^{2}
+(\overline{D_{1}\Phi}-i\overline{D_{2}\Phi})(D_{1}\Phi+iD_{2}\Phi)+F_{12}\right\}\,.
\ee
In deriving (2.2), we have discarded the boundary terms which give vanishing contributions at spatial infinity.
As the first two terms in (2.2) are both non-negative, 
the minimal $E$ is obtained when $A_{i}$ and $\Phi$ satisfy the Bogomolny equations
\be
F_{12}+\frac{1}{2}(|\Phi |^{2}-1)=0\,,
\ee
\be
D_{1}\Phi+iD_{2}\Phi=0 \,.
\ee
The minimal value of $E$, which is related to the winding number through Stokes' theorem, is 
\be
E=\frac{1}{2}\int d^{2}x\, F_{12}=N\pi \,,
\ee
and it can be interpreted as the energy of $N$ non-interacting vortices.

If we introduce the complex coordinate $z=x^{1}+ix^{2}$, (2.3) and (2.4) can be written as
\be
iF_{z\bar{z}}=\frac{1}{4}(|\Phi|^{2}-1)\,,
\ee
\be
D_{\bar{z}}\Phi=\partial_{\bar{z}}\Phi-iA_{\bar{z}}\Phi=0 \,.    
\ee
Equation (2.7) allows us to write $A_{\bar{z}}= -i\partial_{\bar{z}}\log\Phi$. 
We can express $\Phi$ in terms of a gauge invariant
quantity $h$ and a phase factor $\chi$ as $\Phi=e^{\frac{1}{2}h+i\chi}$, 
where the boundary condition for $\Phi$ implies $h\to 0$ at spatial infinity.
Substituting these into (2.6), we obtain the gauge invariant governing equation for the vortex solutions \cite{JT}
\be
4\partial_{z}\partial_{\bar{z}}h-e^{h}+1=4\pi\sum_{r=1}^{N} \delta(z-Z_{r})\,,
\ee
where $\{Z_{1},\dots ,Z_{N}\}$ are the vortex positions in $\mathbbm{C}$. 
These positions are taken as distinct, simple zeros of $\Phi$, although they can coalesce.
There is a unique solution for any choice of positions.
Notice (2.8) has a form very similar to the Liouville equation on a punctured Riemann surface.
The Higgs vacuum expectation value 1 in (2.8) sets the scale for the system; it thus breaks the conformal invariance.

Close to the $r$-th vortex position $Z_{r}$, $h$ has the following expansion
\bea
h&=&\log|z-Z_{r}|^{2}+a_{r}+\frac{1}{2}\bar{b_{r}}(z-Z_{r})+\frac{1}{2}b_{r}(\bar{z}-\bar{Z_{r}}) \nonumber\\
&&+\bar{c_{r}}(z-Z_{r})^{2}-\frac{1}{4}|z-Z_{r}|^{2}+c_{r}(\bar{z}-\bar{Z_{r}})^{2}+\dots ~\,.
\eea
The expansion coefficients $a_{r}$, $b_{r}$, $\bar{b}_{r}$, $c_{r}$ and $\bar{c}_{r}$ 
are all functions of the separations between the  
$r$-th vortex position $Z_{r}$ and all other vortex positions $Z_{s}$, $s\ne r$. 
The coefficient $a_{r}$ plays the role of a local scaling factor. $b_{r}$ and 
$\bar{b}_{r}$ measure the deviation from circular symmetry 
of $h$ around $Z_{r}$ due to interactions with other vortices.
Samols showed that $b_{r}$ and $\bar{b}_{r}$ play a
central role in the formula for the metric on the moduli space. 
He calculated that the metric is \cite{S}
\be
\sum_{r,s=1}^{N}g_{rs}dZ_{r}d\bar{Z}_{s}
=\sum_{r,s=1}^{N}\left(\delta_{rs}+2\frac{\partial b_{s}}{\partial Z_{r}}\right)dZ_{r}d\bar{Z}_{s}\,.
\ee
Notice that while this formula is very general, 
it is not explicit, as we do not have the exact analytic expression for $b_{s}$ in general.
However, we can deduce from the hermiticity of the metric that
\be
\frac{\partial b_{s}}{\partial Z_{r}}=\frac{\partial \bar{b}_{r}}{\partial \bar{Z}_{s}} \,,
\ee
and from this it follows easily that the metric is K\"ahler.     
The translational invariance of the entire system 
implies that $\sum_{r=1}^{N}b_{r}=\sum_{r=1}^{N}\bar{b}_{r}=0$ and the rotational invariance gives 
$\sum_{r=1}^{N}b_{r}\bar{Z_{r}}=\sum_{r=1}^{N}\bar{b}_{r}Z_{r}$ \cite{R}.

In an analysis of the conservation laws of a model of first order vortex dynamics \cite{MN}, 
it has been shown that some conserved quantities can be expressed in terms of integrals involving $h$ and its derivatives. 
The integral of $h$ itself can also be computed. For $N$ non-coincident vortices
\be
\lim_{\epsilon\to 0}\int_{\tilde{\mathbbm{C}}}d^{2}x\,h = -\pi\sum^{N}_{r=1}(b_{r}\bar{Z_{r}}+\bar{b_{r}}Z_{r}+6)\,.
\ee 
The integration region $\tilde{\mathbbm{C}}$ 
is the entire complex plane $\mathbbm{C}$ with the $N$ small disks of radius $\epsilon$ 
centred at the vortex locations $\{Z_{1},\dots ,Z_{N}\}$ being punctured out, 
and the limit $\epsilon\to 0$ taken. 

In this paper, we aim to find a suitable integral expression for the K\"ahler potential $\mathcal{K}$
involving functions of, and derivatives of, the gauge invariant quantity $h$, 
such that the mixed double derivative of $\mathcal{K}$ with 
respect to the holomorphic and anti-holomorphic collective coordinates gives us the K\"ahler metric in the form (2.10).

\section{The Vortex K\"ahler Potential and the Scaling Integral}
\setcounter{equation}{0}

In the K\"ahler metric on the moduli space of vortices, 
the expansion coefficient $b_{r}$ plays the central role. The exact expression for $b_{r}$ is
only known in a few limiting cases, e.g. $N$ overlapping vortices on a sphere \cite{M2}. 
However we know $b_{r}$ obeys the hermiticity identity (2.11) and a recently derived symmetric identity \cite{RS}
\be
\frac{\partial b_{r}}{\partial \bar{Z_{s}}}=\frac{\partial b_{s}}{\partial \bar{Z_{r}}}\,. 
\ee
Clearly (2.11) and (3.1) can both be satisfied 
if there exists a real function $\widetilde{\mathcal{K}}$ of the collective coordinates
$\{Z_{1},\dots ,Z_{N};\bar{Z_{1}},\dots,\bar{Z_{N}}\}$ such that
\be
\frac{\partial \widetilde{\mathcal{K}}}{\partial \bar{Z_{r}}}=2b_{r}
~,~\frac{\partial \widetilde{\mathcal{K}}}{\partial Z_{r}}=2\bar{b_{r}}\,.
\ee
Moreover, the existence of such $\widetilde{\mathcal{K}}$, which plays the role of interacting part for the K\"ahler potential, 
follows from these identities. 
Solving (3.2) for $\widetilde{\mathcal{K}}$ is analogous to the 
so-called $\bar{\partial}$-problem in the mathematical literature.
We do not solve these equations separately, but consider the following linear combination of the equations
\be
\sum_{r=1}^{N}\left\{Z_{r}\frac{\partial}{\partial Z_{r}}
+\bar{Z_{r}}\frac{\partial}{\partial \bar{Z_{r}}}\right\}\widetilde{\mathcal{K}}=
2\sum_{r=1}^{N}\left\{Z_{r}\bar{b_{r}}+\bar{Z_{r}}b_{r}\right\}\,.
\ee
This can be written as 
\be
\sum_{r=1}^{N}\left(\lambda_{r}\frac{\partial}{\partial \lambda_{r}}\right)\widetilde{\mathcal{K}}
=2\sum_{r=1}^{N}\left\{Z_{r}\bar{b_{r}}+\bar{Z_{r}}b_{r}\right\}\,,
\ee
where $Z_{r}=\lambda_{r}e^{i\phi_{r}}$, and $\lambda_{r}$ and $\phi_{r}$ 
are the distance and the angle of the $r$-th vortex away from the origin. 
On one side we have the overall scaling operator of the vortex moduli space acting on $\widetilde{\mathcal{K}}$,
and on the other side a real quantity which is essentially the integral of $h$, using (2.12). 
We have now reduced a problem depending on $2N$ parameters 
$\{Z_{1},\dots ,Z_{N}; \bar{Z_{1}},\dots ,\bar{Z_{N}}\}$ 
to a problem that only depends on the $N$ parameters $\{\lambda_{1},\dots ,\lambda_{N}\}$. 
Effectively we have eliminated the angular dependences 
and shall subsequently keep the angles $\{\phi_{1},\dots ,\phi_{N}\}$ fixed. 

With the scaling operator in mind, we make an ansatz that all $\lambda_{r}$ are parameterised 
by a single variable $\tau$, i.e. $\lambda_{r}\equiv \lambda_{r}(\tau)$ and $Z_{r}=Z_{r}(\tau)=\lambda_{r}(\tau)e^{i\phi_{r}}$, 
where $\tau$ is a dimensionless parameter measuring how $\widetilde{\mathcal{K}}$ 
changes under overall scaling of the moduli space.
So we can express the overall scaling operator in terms of $\tau$ 
\bea
\tau\frac{d}{d\tau}\widetilde{\mathcal{K}}
&=&\sum_{r=1}^{N}\left\{\tau\left(\frac{d Z_{r}(\tau)}{d\tau}\right)\frac{\partial}{\partial Z_{r}}
+\tau\left(\frac{d\bar{Z}_{r}(\tau)}{d\tau}\right)\frac{\partial}{\partial\bar{Z}_{r}}\right\}\widetilde{\mathcal{K}}\nonumber\\
&=&\sum_{r=1}^{N}\left\{Z_{r}\frac{\partial}{\partial Z_{r}}
+\bar{Z_{r}}\frac{\partial}{\partial \bar{Z_{r}}}\right\}\widetilde{\mathcal{K}}\,.
\eea
The second line in (3.5) follows from the first if $\tau\frac{d Z_{r}}{d\tau}=Z_{r}~\forall r$, 
i.e. $Z_{r}$ itself is also proportional to the scaling parameter $\tau$ 
and we have restricted all the vortices to the "scaling" motion. 
We can set the constant of proportionality to be $Z_{r}^{0}$ to obtain $Z_{r}(\tau)=Z_{r}^{0}\tau$,
so that $\{Z_{1}^{0},\dots ,Z_{N}^{0}\}$ are the vortex positions where we want to evaluate the K\"ahler potential.           
Moreover, with this scaling argument, 
we can now combine (3.3) and (3.5) and write an integral expression for $\widetilde{\mathcal{K}}$ 
by integrating with respect to $\tau$. 
We take $\tau$ ranging from $\infty$ to $1$. 
This corresponds to bringing the vortices from spatial infinity 
to their desired positions, so we obtain  
\bea
\widetilde{{\mathcal{K}}}
&=&2\int^{1}_{\infty}\frac{d\tau}{\tau}\,\sum_{r=1}^{N}\left(\bar{b_{r}}
(\tau)Z_{r}(\tau)+b_{r}(\tau)\bar{Z}_{r}(\tau)\right)\nonumber\\
&=&-2\int^{1}_{\infty}\frac{d\tau}{\tau}\,\left\{\frac{1}{\pi}\int d^{2}x\, h({\bf{x}};\tau)+6N\right\}\,,
\eea
where the $\epsilon\to 0$ limit is implied in the integral of $h$. We have 
written $b_{r}\equiv b_{r}(\tau)$ to highlight that $b_{r}$ only depends on $\tau$, 
as we have restricted the vortices to the ``scaling'' motion,
and $h(\bf{x};\tau)$ means the function $h$ in the plane, 
again with this scaling motion of the vortices. 
With a suitable change of coordinates, this expression reduces 
to the expression given in \cite{KL} for the case of $N=2$. 
The entire K\"ahler potential by this scaling argument is therefore given by
\be
{\mathcal{K}}
=\sum_{r=1}^{N}Z_{r}\bar{Z_{r}}
-2\int^{1}_{\infty}\frac{d\tau}{\tau}\,\left\{\frac{1}{\pi}\int d^{2}x\, h({\bf{x}};\tau)+6N\right\}\,,
\ee
where the first term gives the $\delta_{rs}$ in (2.10).

As a simple test for this formula, we consider the case of two well separated vortices. We work in the centre of mass frame, 
and use the asymptotic $b_{r}$ values as calculated in \cite{MS}. 
Using the notation $Z_{1}=\sigma e^{i\theta}=-Z_{2}$
and $b_{1}=b(\sigma)e^{i\theta}=-b_{2}$ where $\sigma$ is the separation from the origin, 
we have $\sum_{r=1}^{2}\left\{Z_{r}\bar{b_{r}}+\bar{Z_{r}}b_{r}\right\}=4\sigma b(\sigma)$. 
$b(\sigma)$ equals $\frac{q^{2}}{2\pi^{2}}K_{1}(2\sigma)$ for large $\sigma$, 
where the constant $q^{2}$ was calculated to be $4\pi^{2}\sqrt{8}$ by Tong using string duality \cite{T}.
Applying the integral expression (3.7), we find the asymptotic K\"ahler potential for two vortices is 
\be
{\mathcal{K}}=2\sigma^{2}-8\sqrt{8}K_{0}(2\sigma)\,.
\ee
The functions $K_{0}, K_{1}$ are the modified Bessel functions of the second kind, 
and we have used Bessel function identities in deriving (3.8). 
Equation (3.8) coincides with the formula for the K\"ahler potential given in \cite{MS} up to an overall factor of $\pi$.

\section{Regularized Liouville Action and Vortex K\"ahler Potential}
\setcounter{equation}{0}

In this section we present a different approach to solving the equations (3.2). 
It turns out that $\widetilde{\mathcal{K}}$ is
in fact a suitably regularized modified Liouville action which gives rise to the vortex equation (2.8). We shall consider 
the case of vortices on $\mathbbm{C}$ here and discuss vortices on the hyperbolic plane ${\mathbbm{H}}^{2}$ in the next section.
The motivation for this section was drawn from the striking similarity of $b_{r}$ to the 
so-called ``accessory parameters'' in the Liouville field theory.
We first review ideas concerning Liouville theory.

Let us consider a Liouville field $\phi$ defined over an $n$-punctured Riemann sphere, 
$\Sigma\cong\hat{\mathbbm{C}}/\{z_{1},z_{2},\dots ,z_{n}\}$, $\hat{\mathbbm{C}}=\mathbbm{C}\cup\{\infty\}$, where $z_{n}=\infty$ and $n\geq 3$. 
In the following, we shall adopt the conventions in \cite{TZ}, \cite{Mat}. $\phi$ satisfies the Liouville equation
\be
\partial_{z}\partial_{\bar{z}}\phi=\frac{1}{2}e^{\phi}\,,
\ee
and we assume the punctures are parabolic singularities, with asymptotic behaviour 
\be
\phi=\cases{-2\log|z-z_{r}|-2\log\left|\log|z-z_{r}|\right|+O(1)~~&\textrm{as $z\to z_{r}$, $r\neq n$}\, ,
    \cr -2\log|z|-2\log\log|z|+O(1)~~&\textrm{as $z\to z_{n}=\infty$}\,.}
\ee
$\phi$ determines the Poincar\'e metric on $\Sigma$, $ds^{2}=e^{\phi}dz d\bar{z}$, 
with the Ricci scalar given by $R_{\Sigma}=-4e^{-\phi}\partial_{z}\partial_{\bar{z}}\phi$. 
The Liouville equation implies $R_{\Sigma}=-2$. (The Gaussian curvature is half this.)
 
The uniformization theorem of the Riemann surface
states that $\Sigma$ has a universal covering space, 
the Poincar\'e half plane $\mathbbm{H}^{2}=\{z\in \mathbbm{C}, \textrm{Im}~z> 0\}$,
and $\Sigma\cong \mathbbm{H}^{2}/\Gamma$. Here $\Gamma$ 
is a finitely generated Fuchsian group, which is a subgroup of $PSL(2,\mathbbm{R})$ acting discretely on $\mathbbm{H^{2}}$.
$e^{\phi}$ can be expressed as 
\be
e^{\phi}=\frac{4\partial_{z}w\overline{\partial_{z}w}}{(1-w\bar{w})^{2}}~,~w=\frac{u_{1}}{u_{2}}\,,
\ee
where $u_{1}$ and $u_{2}$ are a pair of suitably normalized, linearly independent, holomorphic solutions 
of the Fuchsian differential equation with the monodromy group $\Gamma$,
\be
\partial_{z}\partial_{z}u+\frac{1}{2}T_{\phi}(z)u=0\,.
\ee
$T_{\phi}(z)$ is a meromorphic function over $\hat{\mathbbm{C}}$ 
and plays the role of $zz$ component of the energy-momentum tensor for the Liouville field $\phi$,
\be
T_{\phi}(z)=\partial_{z}\partial_{z}\phi-\frac{1}{2}(\partial_{z}\phi)^{2}~~,~~\partial_{\bar{z}}T_{\phi}(z)=0\,.
\ee
Here, $T_{\phi}(z)$ is given by
\bea
T_{\phi}(z)&=&\sum_{r=1}^{n-1}\left\{\frac{1}{2(z-z_{r})^{2}}+\frac{c_{r}}{(z-z_{r})}\right\}\,,\nonumber \\
           &=&\frac{1}{2z^{2}}+\frac{c_{n}}{z^{3}}+O\left(\frac{1}{z^{4}}\right)\,,
\eea
where the second expression is the expansion around $z=\infty$.
The coefficient of each term $1/(z-z_{r})^{2}$, $r=1,\dots,n-1$ and also of $1/z^{2}$ for $z_{n}=\infty$,
is half the conformal weight of each $z_{r}$, 
so in this case each parabolic singularity $z_{r}$ has conformal weight $1$. 
The complex numbers $c_{r}$, $r=1,\dots ,n$, are the accessory parameters.
They are uniquely determined by the positions of the punctures
$z_{1},\dots ,z_{n}$. If we match the first expression with the second expression 
for $T_{\phi}(z)$ as $z\to \infty$, we can derive the following constraints
\be
\sum_{r=1}^{n-1}c_{r}=0~~,~~\sum_{r=1}^{n-1}c_{r}z_{r}=1-\frac{n}{2}
~~,~~\sum_{r=1}^{n-1}(z_{r}+c_{r}z_{r}^{2})=c_{n}\,.
\ee
One notices that the first two of these look remarkably 
similar to the constraints $\sum_{r=1}^{N}\bar{b}_{r}=0$ 
and $\sum_{r=1}^{N}\bar{b_{r}}Z_{r}$ is real, with $\bar{b_{r}}$ analogous to $c_{r}$ !
For the case of $n>3$, these three constraints allow one to express any three accessory parameters, 
say $c_{n-2}$, $c_{n-1}$ and $c_{n}$, in terms of the others.
The other $n-3$ accessory parameters $c_{r}$ are like $b_{r}$, which are in general difficult to calculate explicitly. 
However, based on consideration of the conformal Ward identity, 
the Polyakov conjecture states that, 
for a Riemann sphere with $n$ punctures, the suitably regularized Liouville action 
evaluated at the classical solution is the generating function for $c_{r}$ \cite{P}, i.e.
\be
c_{r}=-\frac{1}{2\pi}\frac{\partial S_{cl}}{\partial z_{r}}~~~~r=1,\dots ,n-3\,.
\ee
The regularized Liouville action is 
\be
S_{cl}
=\lim_{\epsilon\to 0}
\left\{\frac{i}{2}\int_{\Sigma_{\epsilon}}dz\wedge d\bar{z}\,\left(\partial_{z}\phi\partial_{\bar{z}}\phi+e^{\phi}\right)
+2\pi\left(n\log\epsilon+2(n-2)\log|\log\epsilon|\right)\right\}\,,
\ee
where we have used complex coordinates, and $\frac{i}{2}dz\wedge d\bar{z}=dx^{1}\wedge dx^{2}$.
The integration region is 
$\Sigma_{\epsilon}=\mathbbm{C}/(\{\bigcup_{r=1}^{n-1}|z-z_{r}|<\epsilon\}\cup\{|z|>\frac{1}{\epsilon}\})$,
whose boundaries are a circle near
$\infty$ and infinitesimal circles around the punctures. 
The Polyakov conjecture has been proved by Takhtajan and Zograf, 
and the result has been extended to more general singularities \cite{TZ}, \cite{CMS}, \cite{HJ}.

We now turn to the vortex equation and show that one may understand $b_{r}$ in a similar way. 
Let us consider the analogous $zz$ component of the energy-momentum tensor for $h$,
$T_{h}=\partial_{z}\partial_{z}h-\frac{1}{2}(\partial_{z}h)^{2}$. 
Using the expansion of $h$ around each vortex centre (2.9), we have
\be
T_{h}=T_{h}(z,\bar{z})=-\sum_{r=1}^{N}\left\{\frac{3}{2(z-Z_{r})^{2}}+\frac{\bar{b_{r}}}{2(z-Z_{r})}\right\}+O(1)\,.
\ee
The $O(1)$ terms are not all holomorphic.
In the notation of \cite{TZ}, 
$h$ would correspond to a field with the conformal weight $-3$ at each point $Z_{r}$. This differs from
the strictly positive conformal weights in \cite{TZ}, and is caused 
by the opposite asymptotic behaviour of $h$ around the vortex centres $Z_{r}$ 
to the Liouville field $\phi$ around the punctures $z_{r}$, c.f. (2.9) and (4.2). 
If we consider the metric $e^{h}dzd\bar{z}$, we will find that each vortex corresponds to a conical singularity of deficit angle $-2\pi$.
Despite this, we can clearly identify $-\bar{b_{r}}/2$ as an accessory parameter.

So our task now is finding the analogous
regularized action for $h$, as it will be the interacting part of the K\"ahler potential. 
The unregularized action for $h$ is given by
\be
S_{h}=\frac{i}{2\pi}\int_{\tilde{\mathbbm{C}}} 
dz\wedge d\bar{z}\,\left(2\partial_{z}h\partial_{\bar{z}}h+e^h-h-1\right)\,,
\ee
where the integration region $\tilde{\mathbbm{C}}$ is the same as given following (2.12).

As the integrals of $h$ and $e^{h}-1$ are known to be finite as $\epsilon\to 0$, 
the only singular term in the integral comes from $\partial_{z}h\partial_{\bar{z}}h$, 
which gives a contribution of $-4N\log\epsilon+O(1)$. 
In addition to the removal of this $\epsilon$ dependent term, 
we require of a regularized action that when it is evaluated on a classical solution, 
i.e. $h$ satisfying the field equation    
$4\partial_{z}\partial_{\bar{z}}h-e^{h}+1=0$ on $\tilde{\mathbbm{C}}$ with fixed singularities at $\{Z_{1},\dots ,Z_{N}\}$,
it should be stationary against physical variations $\delta h$ of $h$.
Clearly the variation of $S_{h}$ can only come from the boundary contributions, 
as the other terms can be eliminated by using the field equation, so we have
\bea
\delta S_{h}&=&
\frac{i}{2\pi}\int_{\tilde{\mathbbm{C}}}dz\wedge d\bar{z}\,
\left(2\partial_{z}(\delta h \partial_{\bar{z}}h)+2\partial_{\bar{z}}(\delta h\partial_{z} h)\right)
\nonumber\\
&=&-\frac{i}{2\pi}\sum_{r=1}^{N}\left\{2\int_{\gamma_{r}}d\bar{z}\, \delta h \partial_{\bar{z}} h-2\int_{\gamma_{r}}dz\,\delta h \partial_{z}h\right\}\,,
\eea
where $\gamma_{r}$ is the small circle of radius $\epsilon$ centred at $Z_{r}$.
We would like to be as general as possible about the form of $\delta h$. However, with the vortex centres $\{Z_{1},\dots ,Z_{N}\}$ fixed, 
we have to restrict the leading order behaviour
of $h$ around $Z_{r}$ to remain $h=\log|z-Z_{r}|^{2}$, 
otherwise the Higgs field may have an unphysical branch point at $Z_{r}$ if we vary the power dependence. 
So we suppose that near $Z_{r}$,
\be
\delta h=\delta a_{r}+\frac{1}{2}\delta \bar{b}_{r}(z-Z_{r})+\frac{1}{2}\delta b_{r}(\bar{z}-\bar{Z_{r}})+O(\epsilon^{2})\,.
\ee
Substituting (4.13) into (4.12), one can show that $\delta S_{h}=-4\sum_{r=1}^{N}\delta a_{r}$ by a similar calculation to those in \cite{MN}. 
In order to cancel this variation, we need to add a term $4\sum_{r=1}^{N}a_{r}$ to the unregularized action $S_{h}$.
So combined with the $\log\epsilon$ term, we obtain our regularized action for $h$
\be
S_{h}^{reg.}=\lim_{\epsilon \to 0}\left\{
\frac{i}{2\pi}
\int_{\tilde{\mathbbm{C}}}dz\wedge d\bar{z}\,
\left(2\partial_{z}h\partial_{\bar{z}}h+e^h-h-1\right)+4\sum_{r=1}^{N}a_{r}+4N\log\epsilon\right\}\,.
\ee 
This is stationary for solutions of the field equation. 
From now on, we only consider the value of $S_{h}^{reg.}$ evaluated at the classical solutions,
so it becomes a function of $\{Z_{1},\dots ,Z_{N};\bar{Z_{1}},\dots ,\bar{Z_{N}}\}$.

We want to demonstrate that $S_{h}^{reg.}$ is, up to a sign, 
indeed the real function $\widetilde{\mathcal{K}}$ in (3.2) by a direct calculation. Consider 
\bea
\frac{\partial S_{h}^{reg.}}{\partial Z_{s}}
&=&\frac{i}{2\pi}\int_{\tilde{\mathbbm{C}}}
dz\wedge d\bar{z}\,\frac{\partial}{\partial Z_{s}}\left(2\partial_{z}h\partial_{\bar{z}}h+e^h-h-1\right)
+4\sum_{r=1}^{N}\frac{\partial a_{r}}{\partial Z_{s}}\nonumber\\
&~&-\frac{i}{2\pi}\int_{\gamma_{s}}d\bar{z}\left(2\partial_{z}h\partial_{\bar{z}}h+e^h-h-1\right)\,.
\eea
The contribution in the second line of (4.15) 
comes from the fact that as we take the partial derivative $\frac{\partial}{\partial Z_{s}}$, 
we are calculating the change of $S^{reg.}_{h}$ due to the movement of the $s$-th vortex; 
therefore we need to take into account the movement of the small circle $\gamma_{s}$ caused by the movement of the $s$-th vortex,
hence the extra term. One can also use the field equation to derive a useful identity
\be
dz\wedge d\bar{z}\frac{\partial}{\partial Z_{s}}\left(2\partial_{z}h\partial_{\bar{z}}h+e^h-h-1\right)=
d\left\{2d\bar{z}(\partial_{Z_{s}}h\partial_{\bar{z}}h)-2dz(\partial_{Z_{s}}h\partial_{z}h)\right\}\,.
\ee 
Using this, one can rewrite the terms in the first line of (4.15), hence
\bea
\frac{\partial S_{h}^{reg.}}{\partial Z_{s}}
&=&-\frac{i}{2\pi}\sum_{r=1}^{N}
\left\{2\int_{\gamma_{r}}d\bar{z}\,(\partial_{Z_{s}}h\partial_{\bar{z}}h)-2\int_{\gamma_{r}}dz\,(\partial_{Z_{s}}h\partial_{z}h)\right\}
+4\sum_{r=1}^{N}\frac{\partial a_{r}}{\partial Z_{s}}\nonumber\\
&~&-\frac{i}{2\pi}\int_{\gamma_{s}}d\bar{z}\,\left(2\partial_{z}h\partial_{\bar{z}}h+e^h-h-1\right)\,.
\eea
The boundary integrals above can be evaluated using the expansion (2.9), and we find
\bea
\frac{\partial S_{h}^{reg.}}{\partial Z_{s}}&=&
\left(\bar{b_{s}}-2\sum_{r=1}^{N}\frac{\partial{a_{r}}}{\partial Z_{s}}\right)
+\left(2\bar{b_{s}}-2\sum_{r=1}^{N}\frac{\partial a_{r}}{\partial Z_{s}}\right)
+4\sum_{r=1}^{N}\frac{\partial a_{r}}{\partial Z_{s}}-\bar{b_{s}}\nonumber\\
&=&2\bar{b_{s}}\,,
\eea
which is our main result.
As $S_{h}^{reg.}$ is manifestly real, we also have $\partial S_{h}^{reg.}/\partial \bar{Z_{s}}=2b_{s}$. 
Hence $S_{h}^{reg.}$ is indeed the generating function of $b_{s}$ and $\bar{b_{s}}$. 

We can now write down the entire K\"ahler potential for the vortices on $\mathbbm{C}$ in this formulation as
\be
{{\mathcal{K}}}=\sum_{r=1}^{N}Z_{r}\bar{Z_{r}}+S_{h}^{reg.}\,.
\ee
$S_{h}^{reg.}$ acts as the interacting part of the K\"ahler potential on the moduli space of vortices. 
It has the symmetries of the $N$-vortex system, 
namely the overall translational and rotational invariance. Therefore
\bea
&&\sum_{r=1}^{N}\frac{\partial S_{h}^{reg.}}{\partial Z_{r}}=0~~~\textrm{translational invariance,}\\
&&\sum_{r=1}^{N}\left\{Z_{r}\frac{\partial}{\partial Z_{r}}-\bar{Z_{r}}\frac{\partial}{\partial \bar{Z_{r}}}\right\}S^{reg.}_{h}=0
~~~\textrm{rotational invariance.}
\eea
However these two expressions just translate into the aforementioned identities $\sum_{r=1}^{N}\bar{b_{r}}=0$ and
$\sum_{r=1}^{N}\bar{b_{r}}Z_{r}=\sum_{r=1}^{N}b_{r}\bar{Z_{r}}$ respectively. 
Notice that these symmetries restrict the forms of additional
holomorphic or anti-holomorphic functions we could add to our K\"ahler potential.

One final remark about $S_{h}^{reg.}$ is that it can be simplified, because $1-e^{h}$ is twice the magnetic field,
whose integral over the plane is a constant in the $N$-vortex sector. 
The term proportional to $1-e^{h}$ in the integrand of (4.14) can therefore be dropped.
However, this simplification only applies to 
$S_{h}^{reg.}$ evaluated on the classical solutions.

\section{The K\"ahler Potential for Hyperbolic Vortices}
\setcounter{equation}{0}

Vortices on the hyperbolic plane $\mathbbm{H}^{2}$ with Ricci scalar $-1$ were first considered by Witten \cite{W}, 
and an expression for the K\"ahler metric on the moduli space was derived by Strachan \cite{IS}.
One distinct difference between the hyperbolic vortices and the flat vortices is that, 
as shown by Witten and Strachan, 
a Liouville field naturally exists in the hyperbolic case.
We can use this Liouville field instead of the gauge invariant quantity $h$ 
to describe the $N$-vortex system. 
Furthermore, the K\"ahler metric on moduli space can be computed exactly for up to two hyperbolic vortices.    

In this section, we demonstrate how the Liouville field naturally arises and
show that the regularized action for it provides the entire K\"ahler potential for the 
vortices on $\mathbbm{H}^{2}$.
This differs from the flat case where 
we had to add the non-interacting part $\sum_{r=1}^{N}Z_{r}\bar{Z_{r}}$ by hand.

Here are some results for vortices on general Riemann surfaces \cite{MN2}.  
For a Riemann surface, we can always choose coordinates to express its metric $g$ as $g=\Omega^{2}(z,\bar{z})dzd\bar{z}$,
where $\Omega^{2}(z,\bar{z})$ is known as the conformal factor.
The field equation for the $N$-vortex system defined on such a surface is 
\be
4\partial_{z}\partial_{\bar{z}}h-\Omega^{2}(z,\bar{z})\left(e^{h}-1\right)=4\pi\sum_{r=1}^{N}\delta(z-Z_{r})\,.
\ee
Vortices exist as before, with arbitrary locations $Z_{r}$ \cite{B}, 
and the generalized Samols metric on the moduli space of $N$-vortices is
\be
\sum_{r,s=1}^{N}\left\{\Omega^{2}(Z_{r},\bar{Z}_{r})\delta_{rs}+2\frac{\partial b_{s}}{\partial Z_{r}}\right\}dZ_{r}d\bar{Z_{s}}\,.
\ee
The quantities $b_{r}$ are defined as earlier, through the expansion (2.9) of $h$ near the vortex centres, 
the only difference being that in the middle quadratic term, the coefficient $\frac{1}{4}$ 
is replaced by $\frac{1}{4}\Omega^{2}(Z_{r},\bar{Z}_{r})$. 

For the hyperbolic plane $\mathbbm{H}^{2}$ in the standard disk model, with Ricci scalar $R=-1$, 
the conformal factor $\Omega^{2}(z,\bar{z})$ is $8/(1-z\bar{z})^{2}$. 
We can also express $\Omega^{2}(z,\bar{z})$ in terms of a Liouville field $\sigma$,
\be
g=e^{\sigma}dzd\bar{z}=\frac{8dzd\bar{z}}{(1-z\bar{z})^{2}}~~,~~|z|<1\,,
\ee
where $\sigma$ satisfies the Liouville equation $\partial_{z}\partial_{\bar{z}}\sigma=\frac{1}{4}e^{\sigma}$.
One can get back to (4.1) by shifting $\sigma$ to $\sigma+\log 2$, which will correspond to scaling the hyperbolic plane.
But in this section we shall fix the Ricci scalar to be $-1$.

Now consider a conformal transformation on the hyperbolic plane,
$g\to\hat{g}=e^{h}g=e^{h+\sigma}dzd\bar{z}$.   
The Ricci scalar transforms to
\bea
\hat{R}&=&e^{-h}(R-\bigtriangledown^{2}_{g}h) \nonumber\\
           &=&e^{-h}(R-e^{-\sigma}\bigtriangledown^{2}h)\,,
\eea
where $\bigtriangledown_{g}$ is the Laplace-Beltrami operator for $g$ and $\bigtriangledown^{2}=4\partial_{z}\partial_{\bar{z}}$ is the ordinary Laplacian. 
Let us compare this with the equation (5.1) satisfied by $h$ with $\Omega^{2}=e^{\sigma}$.
Away from the singularities, this is 
\be
\bigtriangledown^{2}h-e^{\sigma}(e^{h}-1)=0\,.
\ee
Provided $\hat{R}=R=-1$, we see that equations (5.5) and (5.4) are the same. Moreover,   
the field $\psi=\sigma+h$, 
the conformal factor of the new metric $\hat{g}=e^{\psi}dzd\bar{z}$, also satisfies the Liouville equation 
$\partial_{z}\partial_{\bar{z}}\psi=\frac{1}{4}e^{\psi}$.
 
Hence, we can give a geometrical interpretation to the squared magnitude of the Higgs field $|\Phi|^{2}=e^{h}$ of hyperbolic vortices;
it plays the role of a conformal factor changing the hyperbolic plane $\mathbbm{H}^{2}$ to another 
hyperbolic plane $\hat{\mathbbm{H}}^{2}$ of the same curvature, 
however with conical singularities at $\{Z_{1},\dots,Z_{N}\}$. 
Since $\psi$ is a Liouville field, 
we can, as in (4.3), express $e^{\psi}$ in terms of some meromorphic function $f(z)$. 
$f$ needs to satisfy the condition $|f|<1$ for $|z|<1$, 
to map disk into disk, and also satisfy the boundary condition $|f|\to 1$ as $|z|\to 1$. 
We can then express the metric $\hat{g}$ as follows,  
\be
\hat{g}
=e^{\psi}dzd\bar{z}=\frac{8\left|\frac{df}{dz}\right|^{2}dzd\bar{z}}{(1-f\bar{f})^{2}}
=\frac{8dfd\bar{f}}{(1-f\bar{f})^{2}}\,.
\ee
$f(z)$ can be viewed as a map from $\hat{\mathbbm{H}}^{2}$ to $\mathbbm{H}^{2}$,
and the metric $\hat{g}$ is the pull-back of the standard hyperbolic metric on $\mathbbm{H}^{2}$. 
Conical singularities occur on $\hat{\mathbbm{H}}^{2}$, because the inverse of $f$ is multivalued in general. 
One can also write $\hat{g}$ as
\be
\hat{g}=e^{h}\frac{8dzd\bar{z}}{(1-z\bar{z})^{2}}=
\left|\frac{df}{dz}\right|^{2}\frac{(1-z\bar{z})^{2}}{(1-f\bar{f})^{2}}\frac{8dzd\bar{z}}{(1-z\bar{z})^{2}}\,,  
\ee
and as $\frac{(1-z\bar{z})}{(1-f\bar{f})}$ is real, 
we can choose the phase of $f(z)$ to express the Higgs field $\Phi$ in terms of $f(z)$ as
\be
\Phi=\left(\frac{df}{dz}\right)\frac{(1-z\bar{z})}{(1-f\bar{f})}\,.
\ee
This was the form of $\Phi$ first derived in \cite{IS}.
Clearly, the vortex positions are where $\frac{df}{dz}$ vanishes on the unit disk,
hence for $N$ distinct vortices, the equation $\frac{df}{dz}=0$ should have $N$ distinct roots within the
unit disk. The general form of $f$ satisfying these requirements is given in \cite{W}, \cite{IS}.   
    
We will now follow the procedures of section 4 to derive the K\"ahler potential for the moduli space 
of hyperbolic vortices. In this case, the Liouville field $\psi$ already exists, so we construct the
regularized Liouville action evaluated at a classical solution, and it will turn out to be the entire 
K\"ahler potential.   

First, we write down the expansion of $\psi$ 
around the $r$-th vortex position $Z_{r}$,
\bea
\psi&=&\log|z-Z_{r}|^{2}+A_{r}+\frac{1}{2}\bar{B_{r}}(z-Z_{r})+\frac{1}{2}B_{r}(\bar{z}-\bar{Z_{r}})\nonumber\\
&&+\bar{C_{r}}(z-Z_{r})^{2}+C_{r}(\bar{z}-\bar{Z_{r}})^{2}+\dots~\,.  
\eea
The expansion coefficients of $\psi=h+\log(8/(1-z\bar{z})^{2})$ are related to the ones of $h$ via
\bea
A_{r}&=&a_{r}+\log\left\{\frac{8}{(1-Z_{r}\bar{Z_{r}})^{2}}\right\}\,,\nonumber\\
\bar{B_{r}}&=&\bar{b_{r}}+\frac{4\bar{Z_{r}}}{(1-Z_{r}\bar{Z_{r}})}~,~
B_{r}=b_{r}+\frac{4Z_{r}}{(1-Z_{r}\bar{Z_{r}})}\nonumber\\
\bar{C_{r}}&=&\bar{c_{r}}+\frac{\bar{Z_{r}}^{2}}{(1-Z_{r}\bar{Z_{r}})^{2}}~,~
C_{r}=c_{r}+\frac{Z_{r}^{2}}{(1-Z_{r}\bar{Z_{r}})^{2}}\,.
\eea
One can show by computing the $zz$ component of the energy-momentum tensor 
for $\psi$ that $-\bar{B_{r}}/2$ are the accessory parameters in this
case, and the regularized action of $\psi$ should be the generating function for them. 

The unregularized action for $\psi$ is 
\be
S_{\psi}
=\frac{i}{2\pi}\int_{\mathbbm{D}}dz\wedge d\bar{z}\,\left(2\partial_{z}\psi\partial_{\bar{z}}\psi+e^{\psi}\right)\,,
\ee
where $\mathbbm{D}$ is the unit disk. 
Now we define $\tilde{\mathbbm{D}}=\left\{\mathbbm{D}-\bigcup_{r=1}^{N}|z-Z_{r}|<\epsilon\right\}$,
and using procedures similar to section 4 for calculating
the $\epsilon$ dependent term and applying 
the argument for the stationarity of the action against the physical variation of $\psi$ at the classical solutions, 
we obtain the regularized action for $\psi$ 
\be
S_{\psi}^{reg.}
=\frac{i}{2\pi}\int_{\tilde{\mathbbm{D}}}dz\wedge d\bar{z}\,
\left(2\partial_{z}\psi\partial_{\bar{z}}\psi+e^{\psi}\right)+4\sum_{r=1}^{N}A_{r}+4N\log\epsilon\,,
\ee
where the limit $\epsilon\to 0$ is implied.
We now regard $S^{reg.}_{\psi}$, evaluated on an $N$-vortex solution, as a function just of the vortex positions.
By differentiating $S_{\psi}^{reg.}$, one obtains the following,
\be
\frac{\partial S^{reg.}_{\psi}}{\partial Z_{r}}=2\bar{B_{r}}~~~,~~~\frac{\partial S_{\psi}^{reg.}}{\partial \bar{Z_{r}}}=2B_{r}\,,
\ee
and therefore
\be
\frac{\partial^{2} S_{\psi}^{reg.}}{\partial Z_{s}\partial \bar{Z_{r}}}=2\frac{\partial B_{r}}{\partial Z_{s}}=
\left\{\frac{8\delta_{rs}}{(1-Z_{r}\bar{Z_{r}})^{2}}+2\frac{\partial b_{r}}{\partial Z_{s}}\right\}\,.
\ee
By comparing with (5.2), we see that (5.14) is just the coefficient $g_{rs}$ in the Samols/Strachan metric 
on the moduli space of the vortices on $\mathbbm{H}^{2}$.
This confirms that $S_{\psi}^{reg.}$ is indeed the K\"ahler potential for the hyperbolic vortices.

\section{Instantons on $\mathbbm{R}^{2}\times S^{2}$ and Maciocia's Formula}
\setcounter{equation}{0}

In this section we consider SU(2) Yang-Mills instantons dimensionally reduced on $\mathbbm{R}^{2}\times S^{2}$.
The metric for $\mathbbm{R}^{2}\times S^{2}$ is
\be
ds^{2}=h_{\mu\nu}dx^{\mu}dx^{\nu}=(dx^{1})^{2}+(dx^{2})^{2}+R_{0}^{2}(d\theta^{2}+\sin^{2}\theta~d\varphi^{2})\,.
\ee
The SU(2) Yang-Mills action is given by
\be
S=-\frac{1}{16\pi}\int d^{4}x\sqrt{h}~{\rm{Tr}}(F_{\mu \nu}F^{\mu \nu})\,,
\ee
where $\sqrt{h}\equiv\sqrt{{\rm{det}}\, h_{\mu\nu}}$ and $F_{\mu\nu}=\partial_{\mu}A_{\nu}-\partial_{\nu}A_{\mu}+[A_{\mu},A_{\nu}]$.
The action has a spherical symmetry reflecting the underlying $S^{2}$ geometry. 
Imposing this SO(3) symmetry on the Yang-Mills field \cite{FM}, one finds in a suitable gauge that the gauge potential has components
\bea
&A_{1}^{a}&=(0,0,A_{1}) \nonumber\\
&A_{2}^{a}&=(0,0,A_{2}) \nonumber\\
&A_{\theta}^{a}&=(-\phi_{1},\phi_{2},0) \nonumber\\
&A_{\varphi}^{a}&=(-\phi_{2}\sin\theta, -\phi_{1}\sin\theta,-\cos\theta)\,,
\eea
where $A_{1}$, $A_{2}$, $\phi_{1}$ and $\phi_{2}$ are independent of $\theta$ and $\varphi$. 
After integrating over $S^{2}$, the action is dimensionally reduced to
\be
S=\frac{1}{2}\int d^{2}x\, \left\{\frac{R_{0}^{2}}{2}F_{12}^{2}+\overline{D_{i}\Phi}D_{i}\Phi+\frac{1}{2 R_{0}^{2}}(|\Phi |^{2}-1)^{2}\right\}\,.
\ee
Here $\Phi=\phi_{1}+i\phi_{2}$ and we reobtain the energy (2.1) of the Abelian Higgs model by setting $R_{0}^{2}=2$. 
The self-duality condition for the gauge field on 
$\mathbbm{R}^{2}\times S^{2}$ reduces to the Bogomolny equations for the Abelian Higgs model. 
Therefore the static vortex solutions of winding number $N$ arise as spherically symmetric instantons of charge $N$.

Maciocia presented a formula for the so-called ``second moment'' of the action density of 
self-dual gauge fields in $\mathbbm{R}^{4}$ \cite{AM},
\be
m_{2}=-\frac{1}{16\pi^{2}}\int_{\mathbbm{R}^{4}}|x|^{2}{\rm{Tr}}(F\wedge F)\,.
\ee
$m_{2}$ turned out to be the K\"ahler potential for the natural K\"ahler two-form $\omega$ on the instanton moduli space, 
so that $\omega\sim \partial\bar{\partial}m_{2}$.
As pointed out by the authors of \cite{DHKM}, 
the $N$-instanton moduli space is an $8N$-dimensional hyperK\"ahler space for the SU(2) gauge group.  
We can pick one of the three complex structures of the moduli space and then impose an extra restriction on the gauge field
\be
\frac{\partial^{2} A_{\mu}}{\partial Z_{r}\partial \bar{Z_{s}}}=0~,~\mu=1,2,3,4\,,
\ee
where $\{Z_{1},\dots,Z_{4N};\bar{Z_{1}},\dots ,\bar{Z}_{4N}\}$ are the complex collective coordinates for this complex structure.
This restriction turns out to be crucial for showing that $m_{2}$, upon differentiation, 
gives the standard instanton moduli space metric, associated with what would be the kinetic energy
of the instantons in $(4+1)$-dimensions.
 
Maciocia also made the interesting remark that for a general K\"ahler 4-manifold $X$, with K\"ahler potential $W$, 
the K\"ahler potential $\Psi$ on the moduli space of the instantons 
defined over $X$ can be expressed up to a constant of proportionality as  
\be
\Psi\sim -\int_{X}W~{\rm{Tr}}(F\wedge F)\,. 
\ee
Thus the K\"ahler form on the moduli space of $N$ instantons over $X$ is given by
\be
\omega\sim\partial\bar{\partial}\Psi
=\sum_{r,s}\frac{\partial^{2}\Psi}{\partial Z_{r}\partial \bar{Z_{s}}}dZ_{r}\wedge d\bar{Z_{s}}\,.
\ee

Inspired by Maciocia's remark, the idea in this section is straightforward. 
As instantons on $\mathbbm{R}^{2}\times S^{2}$ 
can be dimensionally reduced to the Abelian Higgs vortices on $\mathbbm{R}^{2}$, 
Maciocia's formula for $\Psi$ should be naturally related to the K\"ahler potential for the 
moduli space of such vortices. 
We just need to restrict to the SO(3) invariant instantons,
which form a complex submanifold of the space of all instantons.
The K\"ahler potential on the base manifold $\mathbbm{R}^{2}\times S^{2}$ with $R_{0}^{2}=2$ is
\be
W=z\bar{z}-8\log\left(\frac{1+\cos\theta}{2}\right)\,.   
\ee
Applying Maciocia's integral formula to the dimensionally reduced action (6.4) and choosing a suitable constant of proportionality, we obtain 
\bea 
\Psi
&=&\frac{1}{8\pi^{2}}\int d^{2}x\int \sin\theta d\theta d\varphi\,W
\left\{F_{12}^{2}+\overline{D_{i}\Phi}D_{i}\Phi+\frac{1}{4}\left(|\Phi |^{2}-1\right)^{2}\right\}\nonumber \\
&=&\frac{1}{2\pi}\int d^{2}x\, (z\bar{z}+8)\left\{F_{12}^{2}+\overline{D_{i}\Phi}D_{i}\Phi+\frac{1}{4}\left(|\Phi |^{2}-1\right)^{2}\right\}\,.
\eea
The integral over $S^{2}$ is finite, despite the logarithmic divergence of $W$ at $\theta=\pi$.
For fields satisfying the Bogomolny equation, we can express the integral in terms of $h$ as
\bea
\Psi&=&\frac{i}{2\pi}\int_{\tilde{\mathbbm{C}}} dz\wedge d\bar{z}\,
\left\{z\bar{z}\partial_{z}\partial_{\bar{z}}(e^{h}-h-1)\right\}+8N \nonumber\\ 
&=&\sum_{p=1}^{N}(Z_{p}\bar{Z_{p}}+\bar{b_{p}}Z_{p}+b_{p}\bar{Z}_{p})+10N\,.
\eea
In evaluating the integral in (6.11), we have used a result in \cite{MN}, 
as the integral is proportional to the total angular momentum calculated there for 
first order vortex dynamics.
If we now differentiate $\Psi$ with respect to $\bar{Z_{s}}$ and then $Z_{r}$, we find
\bea
\frac{\partial^{2}\Psi}{\partial Z_{r} \partial \bar{Z_{s}}}
&=&\left(\delta_{rs}+\frac{\partial b_{s}}{\partial Z_{r}}+\frac{\partial \bar{b_{r}}}{\partial \bar{Z_{s}}}\right)+
\sum_{p=1}^{N}
\left(\frac{\partial^{2}b_{p}}{\partial Z_{r} \partial 
\bar{Z_{s}}}\bar{Z_{p}}+\frac{\partial^{2}\bar{b_{p}}}{\partial Z_{r} \partial \bar{Z_{s}}}Z_{p}\right)
\nonumber \\
&=&g_{rs}+\sum_{p=1}^{N}\left(Z_{p}\frac{\partial}{\partial Z_{p}}+\bar{Z_{p}}\frac{\partial}{\partial \bar{Z_{p}}}\right)
\left(\frac{\partial b_{s}}{\partial Z_{r}}\right)\,, 
\eea
where $g_{rs}$ is the Samols metric in (2.10). We have used the hermiticity identity (2.11) to obtain the second line of (6.12).
One notices that the overall scaling operator of the moduli space,
$\sum_{p=1}^{N}(Z_{p}\frac{\partial}{\partial Z_{p}}+\bar{Z_{p}}\frac{\partial}{\partial \bar{Z_{p}}})$, 
is acting on the interacting part of $g_{rs}$, the quantity 
$\frac{\partial b_{s}}{\partial Z_{r}}$. 
The last quantity depends on the separations between the vortices, 
and therefore is not invariant under such a scaling operation, except for $N=1$, 
where the metric is trivial. 
Hence in general, the second term in (6.12) is non-vanishing. 
This indicates that Maciocia's formula does not provide the K\"ahler potential for the Samols metric,
although it is quite close.

A possible reason why we can not properly apply Maciocia's formula lies in (6.6).
Because the base space for vortices is K\"ahler, and the vortex moduli space is also in general K\"ahler, not hyperK\"ahler,
we do not have the algebra of complex structures as in the hyperK\"ahler case, 
and we do not see how one can necessarily impose 
an analogous condition to (6.6) on the two-dimensional gauge field.    
Let us write down the $U(1)$ gauge potential
$A_{\bar{z}}$ in terms of $h$ and $\chi$. 
From (2.7), $A_{\bar{z}}=-i\partial_{\bar{z}}\log\Phi=-\frac{i}{2}\partial_{\bar{z}}h+\partial_{\bar{z}}\chi$.
The phase factor $\chi$ for the Higgs field $\Phi$ needs to increase 
by $2\pi$ along a closed loop around each (simple) vortex centre.   
A nice gauge choice is $\chi=\frac{1}{2i}\sum_{r=1}^{N}\log\frac{(z-Z_{r})}{(\bar{z}-\bar{Z}_{r})}$, 
which ensures that $A_{\bar{z}}$ is a smooth function globally.
Then, near the $p$-th vortex position $Z_{p}$ 
the double derivative of the gauge potential with respect to the collective coordinates is   
\be
\frac{\partial^{2}A_{\bar{z}}}{\partial Z_{r}\partial \bar{Z_{s}}}=
\frac{\partial}{\partial Z_{r}}\frac{\partial}{\partial \bar{Z_{s}}}
\left(\frac{-i}{2}\partial_{\bar{z}}h+\partial_{\bar{z}}\chi\right)=
\frac{-i}{4}\frac{\partial^{2}b_{p}}{\partial Z_{r}\partial \bar{Z}_{s}}+O(\epsilon)\,.
\ee
We have used the expansion (2.9), and noted that the mixed double derivative of $\partial_{\bar{z}}\chi$ gives zero.
We know from the asymptotic analysis in \cite{MS} that $b_{p}$ is not the sum of a holomorphic and anti-holomorphic function 
of the collective coordinates, so the mixed double derivative of $b_{p}$ in (6.13) does not vanish.
Therefore, in this gauge, the mixed double derivative of $A_{\bar{z}}$ on the left hand side is non-zero. 
Whether it is non-zero in all gauges, we do not know.

In summary, use of Maciocia's formula, combined with dimensional reduction on a suitable manifold,
has got us quite close to the K\"ahler potential, and 
provides us with a possible clue how 
the moduli space metrics of two-dimensional vortices and four-dimensional instantons can be related.

\section{Conclusion}
\setcounter{equation}{0}

In this paper, we have presented three possible approaches to 
calculating the K\"ahler potential for $N$ distinct Abelian Higgs vortices on $\mathbbm{R}^{2}$, 
all of them involving various integrals of the gauge invariant quantity $h=\log|\Phi|^{2}$. 
For vortices on the hyperbolic plane $\mathbbm{H}^{2}$, 
instead of $h$, we calculated the K\"ahler potential in terms of an integral involving the Liouville field $\psi$. 

The first approach uses a scaling argument and appears to be the easiest.
Our result agrees with the K\"ahler potential for two well separated  vortices given in \cite{MS}. 
When vortices are very close to each other, singularities might occur in the
K\"ahler potential, and this approach breaks down,
as one initial assumption in this approach was that the vortices never coincide. 
 
The second approach described in sections 4 and 5 exploits the striking similarities 
between the expansion coefficient $\bar{b_{r}}$ 
and the accessory parameter $c_{r}$ in the Liouville field theory.

In section 4, for vortices on $\mathbbm{R}^{2}$, the regularized action for the field $h$ evaluated on classical solutions 
was shown to be the interacting part of the K\"ahler potential. 
We however need to add the non-interacting part $\sum_{r=1}^{N}Z_{r}\bar{Z}_{r}$ 
by hand to obtain the full K\"ahler potential on the moduli space.   

In section 5, for vortices on the hyperbolic plane $\mathbbm{H}^{2}$ with Ricci scalar $-1$, 
we derived exactly analogous quantities to the accessory parameters. 
The Polyakov conjecture was then used to show that the regularized action for the Liouville field $\psi$ is  
the entire K\"ahler potential on the moduli space of  vortices on $\mathbbm{H}^{2}$.
Our results indicate that the two-dimensional quantum gravity literature 
provides a largely untapped resource for studies of vortices on general Riemann 
surfaces. 

As discussed in \cite{TZ}, \cite{Mat}, in the context of uniformization of a punctured Riemann sphere $\Sigma$, the regularized Liouville action 
for a Liouville field  $\phi$ defined on $\Sigma$ is the K\"ahler potential 
for the so-called Weil-Petersson metric on the complex space $\mathcal{Z}_{n}$, 
where
\be
{\mathcal{Z}}_{n}
=\left\{(z_{1},\dots,z_{n-3})\in 
\mathbbm{C}^{n-3}\mid z_{i}\neq0,1\,\, \textrm{and}\,\, z_{i}\neq z_{k}\, \forall\, i\neq k\right\}\,.
\ee 
On $\Sigma$, we can use suitable M\"obius transformations to map three of the $n$ punctures to $0,1$ and $\infty$,
therefore we can regard the set $\{z_{1},\dots,z_{n-3}\}$ as the remaining $n-3$ puncture coordinates. 
The relation between $\mathcal{Z}_{n}$ and the moduli space $\mathcal{M}_{n}$ of an $n$-punctured Riemann sphere $\Sigma$ is
$\mathcal{M}_{n}\equiv \mathcal{Z}_{n}/S_{n-3}$, where $S_{n-3}$ is the group permuting $\{z_{1},\dots, z_{n-3}\}$.
 
The results in section 4 and 5 are derived using the Polyakov conjecture 
and identifying the vortex locations with the punctures. 
Because of the relation between $\mathcal{M}_{n}$ and $\mathcal{Z}_{n}$, and the fact that the punctured sphere $\Sigma$ 
has a hyperbolic plane $\mathbbm{H}^{2}$ as its universal cover,
it is natural to ask if we can establish a more concrete relation between 
the projected Weil-Petersson metric on $\mathcal{M}_{n}$ for $\Sigma$ and the K\"ahler
metric on the moduli space of the vortices defined on its universal cover $\mathbbm{H}^{2}$.           
If such a question can be answered, we should then try to relate the moduli space of a punctured general Riemann surface, 
and the moduli space of the vortices defined on the covering space of such a Riemann surface. 

The third approach was to consider the Yang-Mills instantons 
on $\mathbbm{R}^{2}\times S^{2}$ which under dimensional reduction yields the vortices on $\mathbbm{R}^{2}$.
Maciocia's formula for the K\"ahler potential on the moduli space of instantons 
was then applied in an attempt to calculate the K\"ahler potential for the vortices.
However, this approach did not succeed in producing the correct K\"ahler potential on the moduli space.
A possible reason for this shortcoming was outlined in section 6. \newline

\noindent{\bf Acknowledgements}\\
H.-Y.C. would like to thank St.~John's College, Cambridge for generous financial support through a Benefactors' Scholarship.
N.S.M. would like to thank L. Cantini, T. Fokas and S. Krusch for valuable discussions.

\end{document}